\documentclass[a4paper,11pt]{article}
\usepackage[toc,page]{appendix}
\usepackage{appendix}
\usepackage{graphicx}
\usepackage{colortbl}
\usepackage{amsmath}
\usepackage{subfigure}
\usepackage{mathtools}
\usepackage[utf8]{inputenc}
\usepackage{float}
\usepackage[normalem]{ulem}
\usepackage{epstopdf}
\usepackage{amsfonts,amsmath,amssymb}
\usepackage{hyperref}

\usepackage{epsfig,multicol,bbm}
\usepackage{color}
\usepackage[dvipsnames]{xcolor}
\usepackage{cite}

\definecolor{darkblue}{rgb}{0.0, 0.0, 0.55}
\definecolor{grey}{rgb}{0.57, 0.64, 0.69}
\definecolor{lightbrown}{rgb}{0.71, 0.4, 0.11}
\newcommand{\tcb}{\textcolor{blue}}

\newcommand{\tcr}{\textcolor{red}}
\newcommand{\tcg}{\textcolor{green}}

\newcommand{\tcgr}{\textcolor{grey}}
\newcommand{\tcbr}{\textcolor{brown}}

\usepackage[most]{tcolorbox}

\newcommand{\be}{\begin{equation}}
\newcommand{\ee}{\end{equation}}

\date{}

\newcommand\fverb{\setbox\pippobox=\hbox\bgroup\verb}
\newcommand\fverbit{\egroup\item[\fbox{\unhbox\pippobox}]}

\newbox\pippobox

\begin{document}

\title{Superradiance of Charged Static Black Hole in Cubic Gravity}
\author{S. N. Sajadi\thanks{Electronic address: naseh.sajadi@gmail.com}\,,\,
Supakchai Ponglertsakul\thanks{Electronic address: supakchai.p@gmail.com}\,,\,
\\
\small Strong Gravity Group, Department of Physics, Faculty of Science, Silpakorn University,\\ \small Nakhon Pathom 73000, Thailand\\
}

\maketitle
\begin{abstract}
Higher curvature gravity usually has complicated field equations, and solving them analytically is strenuous. In this work, we obtain an analytical charged black hole (BH) solution in higher curvature gravity using the thermodynamics of black holes and employing the continued fraction expansion. We investigate the thermodynamics of static black holes using the first law of thermodynamics and the Smarr formula in their proper form for Einstein-cubic gravity (ECG). Next, we obtain the thermodynamic quantities and show that our results are similar to those from solving the field equations. Then, we study the superradiance of the black hole using massless charged-scalar perturbations. We derive the superradiant conditions and compute the amplification factor through direct integration. We demonstrate how the amplification factor will change as a function of the black hole charge and frequency of the incident wave. {We also show that a black hole mass and charge are decreasing in the superradiance region. Finally, we discuss superradiance as a consequence of black hole thermodynamics in ECG.}
\end{abstract}

\section{Introduction}
To study the physics of the universe in different scales, higher curvature gravity theories have attracted considerable attention. Higher curvature gravity refers to extensions of Einstein's gravity that include terms with higher powers of curvature in the action. Such terms can be seen in the renormalization of quantum field theory in curved spacetime and the construction of low-energy effective actions for string theory. One motivation for studying a classical higher curvature theory is to explore black holes. It is interesting to explore which properties of black holes deviate from Einstein's gravity and which are robust features of all higher curvature terms \cite{Myers:1998gt}. However, obtaining the black hole solution for a generic theory when the higher curvature term is considered as an effective theory i.e., treating higher curvature terms perturbatively, is an easy task. The solution obtained from this treatment can be considered as an Einstein solution with some correction terms from the higher curvature gravity. When the higher curvature contribution is non-perturbative, solving the field equations analytically is proving to be difficult. Therefore, most of the black hole solutions are numerical.  In this case, one obtains the BH solutions which are not a solution of pure Einstein gravity but rather a solution of Einstein with higher curvature gravity.

In this paper, we attempt to obtain an approximate analytical black hole solution to a higher curvature gravity when higher curvature terms are considered non-perturbatively. To do so, we consider field equations near the horizon and at the asymptotic region. Then, the black hole solutions are obtained in these respective regions. By using the continued fraction expansion, we obtain a complete non-singular solution exterior to the BH's horizon for a generic coupling constant. However, in the coefficients of the continued fraction expansion, there are two unknown constants that need to be determined. They are determined using the thermodynamics of black holes. In \cite{Sajadi:2020axg} and \cite{Sajadi:2022ybs}, using this approach, we obtain the black hole solutions for Einstein-Weyl square gravity \cite{Stelle:1976gc, Lu:2015cqa} and Einstein-cubic gravity \cite{Bueno:2016xff, Bueno:2016lrh, Hennigar:2018hza, Cisterna:2018tgx, Ahmed:2017jod, Poshteh:2018wqy, Frassino:2020zuv, Feng:2017tev, Jiang:2019kks, Edelstein:2022xlb, Bueno:2018yzo, Bueno:2020uxs, Bueno:2018xqc, Hennigar:2016gkm}, respectively. 

For the Einstein-Weyl square gravity, we discover a branch of BH solutions that is completely different from the Schwarzschild solution. For the Einstein-cubic gravity, the black hole solutions at small radii are found to be different from Schwarzschild BH\cite{Sajadi:2022ybs}.
By following the approach in \cite{Hajian:2023bhq}, in this work, we treat the term corresponding to the coupling constant as thermodynamics quantity. Thus, we modify the first law of thermodynamics and the Smarr formula in accordance with the presence of the coupling constant of the theory. In addition, in this work, we consider the thermodynamics system of a more generic setting, i.e. charged-AdS black hole.

Moreover, we shall consider an interesting phenomenon called superradiance, the field$-$theoretic analog of the Penrose process which refers to the phenomenon that the energy of a test bosonic field (scalar, electromagnetic, or gravitational waves) can be amplified and reflected by the black hole when the incident field on a black hole meets certain conditions \cite{Brito:2015oca}. Moreover, the second law of black hole thermodynamics implies that waves can extract energy from black holes if the superradiant condition is fulfilled \cite{Brito:2015oca}. There are plenty of works on scattering by static singular and regular black holes, including the scattering with massless and massive fields \cite{Jung:2004yh, Doran:2005vm, Dolan:2006vj, Crispino:2009xt, Crispino:2009ki, Baake:2016oku, Balakumar:2020gli, Benone:2015bst, dePaula:2024xnd}. If the scattered field accumulates outside of black hole, superradiance may render black hole spacetime unstable against small perturbations. This phenomena becomes known as superradiant instability. For a charge black hole in anti de Sitter (AdS) and in a cavity, superradiant instability are reported in \cite{Guo:2023ivz,Hod:2012wmy,Degollado:2013bha,Herdeiro:2013pia}. The end-points of this instability is found to be a formation of hairy black hole \cite{Sanchis-Gual:2016tcm,Sanchis-Gual:2015lje,Bosch:2016vcp,Dolan:2015dha,Ponglertsakul:2016wae,Ponglertsakul:2016anb}. For a rotating black hole, superradiance are also studied in different dimensions \cite{Konewko:2023gbu,Dappiaggi:2017pbe,Yang:2022uze}. We refer the interested readers to \cite{Brito:2015oca} for a nice review on this subject. 

In this paper, using the thermodynamics of BH in higher curvature gravity, and motivated by the superradiance of a charged static black hole, we explore the superradiance of charged singular static BH in the framework of cubic gravity. The paper is organized as follows: in section \ref{sec2}, we present the basic formalism of the Einstein-cubic gravity. The special cases for Einstein-cubic gravity are studied in section \ref{sec3}. In section \ref{sec4}, we study the superradiance of the black hole solutions using a charged-scalar field. In section \ref{con}, we summarize our findings.

\section{Basic formalism}\label{sec2}

The general action one may consider in four dimensions is
\begin{equation}
\mathcal{S}=\dfrac{1}{\kappa^2}\int d^{4}x \sqrt{-g}\mathcal{L}+\dfrac{1}{\kappa^2}\int d^{4}x \sqrt{-g}\mathcal{L}_{m},
\end{equation}
where $\kappa$ is the gravitational coupling constant. We particularly consider the gravity part of the Lagrangian as follows
\begin{equation}
\mathcal{L}=\mathcal{R}-2\Lambda +\alpha \mathcal{L}_{1},
\end{equation}
where $\Lambda$ is the cosmological constant, $\mathcal{L}_{1}$ is the Lagrangian of higher curvature gravity and $\mathcal{L}_{m}$ is the Lagrangian of the matter fields. By variation of the action with respect to the metric tensor, one can obtain the corresponding equation of motion as follows
\begin{align}
\mathcal{E}_{ab}=&\mathcal{T}_{ab},
\end{align}
where
\begin{align}
\mathcal{E}_{ab}&=\dfrac{1}{\sqrt{-g}}\dfrac{\delta(\sqrt{-g}\mathcal{L})}{\delta g^{ab}}=\mathcal{P}_{acde}\mathcal{R}_{b}{}^{cde}-\dfrac{1}{2}g_{ab}\mathcal{L}-2\nabla^{c}\nabla^{d}\mathcal{P}_{acdb}, \label{LHS} \\
\mathcal{T}_{ab}&=-\dfrac{2}{\sqrt{-g}}\dfrac{\delta(\sqrt{-g}\mathcal{L}_{m})}{\delta g^{ab}}.
\end{align}
and $ \mathcal{P}^{abcd}=\dfrac{\partial\mathcal{L}}{\partial \mathcal{R}_{abcd}}$. The energy-momentum tensor can be found once the matter Lagrangian $\mathcal{L}_m$ is specified. Now, we consider a static, spherically symmetric line element of a black hole with two metric functions as follows
\begin{equation}
ds^2=-f(r)dt^2+\dfrac{dr^2}{h(r)}+r^2d\theta^2+r^2\sin^2 \theta d\phi^2 .
\end{equation}
With the two unknown metric functions $f(r)$ and $h(r)$, the left-hand side of the field equation \eqref{LHS} yields a very complicated higher-order differential equation (even in a vacuum). Therefore, it is a strenuous task to obtain the metric functions analytically. However, one can find approximate forms of these functions. To do so, we start by assuming a black hole has at least one horizon. Thus, we expand the metric functions around the event horizon as follows:
\begin{align}
f(r)&=\sum_{i=1} f_{i}(r-r_{+})^{i}=f_{1}(r-r_{+})+f_{2}(r-r_{+})^2+f_{3}(r-r_{+})^3+\cdots ,\\
h(r)&=\sum_{i=1} h_{i}(r-r_{+})^{i}=h_{1}(r-r_{+})+h_{2}(r-r_{+})^2+h_{3}(r-r_{+})^3+\cdots .
\end{align}
We can read the temperature from the near horizon expansion as follows
\begin{equation}\label{eqTe}
T=-\left. \dfrac{\kappa}{2\pi}\right\vert_{r_{+}}=\dfrac{\sqrt{f_{1}h_{1}}}{4\pi},\;\;\;\;\;\;\;\;\;\;\kappa^2=-\dfrac{1}{2}\nabla_{\alpha}\xi_{\beta}\nabla^{\alpha}\xi^{\beta},
\end{equation}
here, $\xi$ is timelike Killing vector and $\kappa$ is the surface gravity. Also, we can obtain the thermal entropy using the Wald formula as follows \cite{Wald1, Wald2}
\begin{equation}\label{eqS}
S=-2\pi \int d^{2}x\sqrt{\eta}\dfrac{\delta \mathcal{L}}{\delta \mathcal{R}_{abcd}}\epsilon_{ab}\epsilon_{cd}.
\end{equation}
The other thermodynamics potential corresponds to the cosmological constant, coupling of theory and other fields can be computed via
\begin{equation}\label{eqCh}
\psi_{i}=\int \xi_{H} A_{i},
\end{equation}
here, $A_{i}$ is a gauge field, and $\xi_{H}$ is a Killing vector on the horizon of the black holes.
Therefore, by using the first law of thermodynamics and the Smarr formula we have
\begin{align}
dM=&TdS+\psi_{i}d\alpha_{i},\\M=&2TS+\alpha_{i} \psi_{i}.\label{Smarr}
\end{align}
By substituting \eqref{eqTe}, \eqref{eqS}, and \eqref{eqCh} into \eqref{Smarr}, we can determine the mass of the black hole. Then, by inserting the mass into the first law, one can obtain the partial differential equations for $f_{1}$ and $h_{1}$.
To solve the partial differential equations obtained from the first law of thermodynamics, we have to make assumptions about the relations between $f_{1}$ and $h_{1}$ \cite{Bonanno:2019rsq}. As it has been shown in \cite{Bonanno:2019rsq} for Einstein-Quadratic gravity, the results of $f_{1}\neq h_{1}$ are different from the $f_{1}=h_{1}$. {Finally, by solving the differential equations for $f_{1}$ or $h_{1}$ and inserting them into the equations \eqref{eqTe}, \eqref{eqS} and \eqref{eqCh}, the thermodynamics of the black holes is determined.} In general, this method provides a way to obtain non-Einstein black hole solutions of higher-curvature theories of gravity.
Here, it is worth mentioning that by taking the metric function $f$ as an equipotential surface $f(r)=constant$, the first law of thermodynamics could be derived. The above method is the reverse of this process. 
In the following, we apply this method to Einstein-cubic gravity, a generalized quasi-topological gravity with one metric function.

\subsection{Thermodynamics of Einstein cubic gravity}\label{sec3}

In four dimensions, Einstein Cubic Maxwell theory with the presence of cosmological constant $\Lambda$ is determined by the action \cite{Bueno:2016xff, Bueno:2016lrh, Hennigar:2018hza, Cisterna:2018tgx, Ahmed:2017jod, Poshteh:2018wqy}
\begin{equation}\label{eq1}
\mathcal{S}=\dfrac{1}{16\pi G}\int d^{4}x\sqrt{-g}\left(\mathcal{R}-2\Lambda +\alpha \mathcal{P}-\dfrac{1}{4}\mathcal{F}_{\mu\nu}\mathcal{F}^{\mu\nu}\right),
\end{equation}
where $\mathcal{R}$ represents the Ricci scalar, $\mathcal{F}_{\mu\nu}=\partial_{\mu}A_{\nu}-\partial_{\nu}A_{\mu}$ is the electromagnetic tensor with four potential $A_{\mu}=\left(q/r,0,0,0\right)$, $\alpha$ is coupling constant of the theory, and $\mathcal{P}$ cubic-in-curvature correction to the Einstein-Hilbert action is given as \cite{Bueno:2016xff, Bueno:2016lrh, Hennigar:2018hza, Cisterna:2018tgx, Ahmed:2017jod, Poshteh:2018wqy}
\begin{equation}
\mathcal{P}=12\mathcal{R}_{a}{}^{c}{}_{b}{}^{d}\mathcal{R}_{c}{}^{e}{}_{d}{}^{f}\mathcal{R}_{e}{}^{a}{}_{f}{}^{b}+\mathcal{R}_{a b}{}^{c d}\mathcal{R}_{c d}{}^{e f}\mathcal{R}_{e f}{}^{a b}-12\mathcal{R}_{abcd}\mathcal{R}^{ac}\mathcal{R}^{bd}+8\mathcal{R}_{a}{}^{b}\mathcal{R}_{b}{}^{c}\mathcal{R}_{c}{}^{a}.
\end{equation}
The correction term in four dimensions is dynamical and is not topological or trivial \cite{Bueno:2016xff, Bueno:2016lrh, Hennigar:2018hza, Cisterna:2018tgx, Ahmed:2017jod, Poshteh:2018wqy}.
In this work, we consider a static spherically symmetric line element for describing the geometry of spacetime
\begin{equation}\label{metform}
ds^{2}=-f(r)dt^{2}+\dfrac{dr^{2}}{h(r)}+r^2\left(d\theta^{2}+\sin^{2}\theta d\phi^{2}\right).
\end{equation}
In general, static spherically symmetric metrics metrics do not need to obey $g_{tt}g_{rr} = -1$ or $h=f$ necessarily. However, the field equation of Einstein Cubic gravity admits a solution that obeys the mentioned relation \cite{Bueno:2016xff, Bueno:2016lrh, Hennigar:2018hza, Cisterna:2018tgx, Ahmed:2017jod, Poshteh:2018wqy}. Thus, we shall restrict our consideration only to the case when $h=f$.
Since the metric function is assumed to be non-asymptotically flat (with the presence of $\Lambda$), at the large $r$, we assume that the metric function can be written in the form
\begin{align}\label{eqqqhfo}
 f_{p}(r)=\Lambda_{eff}r^2+&\sum_{n=0}\dfrac{F_{n}}{r^{n}}=\Lambda_{eff}r^2+F_{0}+\dfrac{F_{1}}{r}+\dfrac{F_{2}}{r^2}+....,
\end{align}
where expansion coefficients $F_0, F_1...$ can be determined by inserting the above expansions into the field equations and solving order by order. One obtains, at the large $r$
\begin{align}\label{eqasymp}
f(r)=1+\Lambda_{eff}r^2-&\dfrac{2M}{r}+\dfrac{2q^2}{(1+48\alpha\Lambda_{eff}^2)r^2}+\dfrac{14M^{2}}{\Lambda_{eff}r^4}-\dfrac{4992\alpha q^2\Lambda_{eff}M}{(1+48\alpha \Lambda_{eff}^{2})(-1+96\alpha \Lambda_{eff}^2)r^{5}}\nonumber\\
&+\dfrac{12\alpha(176q^{4}\Lambda_{eff}+41472 M^{2}\alpha^2 F_{2}^{4}+3456\alpha M^{2}\Lambda_{eff}^2+36M^{2})}{(1+48\alpha\Lambda_{eff}^2)^2(48\alpha\Lambda_{eff}^2-1)r^6}+\mathcal{O}(r^{-7}).
\end{align}
The $\Lambda_{eff}$ is found by solving the following relation
\begin{equation}
48\alpha\Lambda_{eff}^3+9\Lambda_{eff}+2\Lambda=0 .
\end{equation}
We also assume that the solution has at least one event horizon $r_+$. Therefore, we expand the function $f(r)$ around the event horizon $ r_{+} $ as
\begin{align}\label{eq7}
f(r)  &= f_{1}(r-r_{+})+f_{2}(r-r_{+})^{2}+f_{3}(r-r_{+})^{3}+...
\end{align}
here
\begin{align}\label{eq9}
{f_{2}}=&\dfrac{2r_{+}^4\Lambda -2r_{+}^2-12\alpha f_{1}^{2}+3r_{+}^3f_{1}+2q^2}{r_{+}^4},\nonumber\\
f_{3}=&-\dfrac{1}{3r_{+}^7(r_{+}^3+48\alpha f_{1}+24\alpha r_{+}f_{1}^2)}\Big(768\alpha f_{1}^{3}r_{+}^6+576\alpha^2 r_{+}^2f_{1}^{3}-4608\alpha^2 f_{1}^4r_{+}^{3}+192\alpha f_{1}q^4-\nonumber\\
&2304\alpha^2 f_{1}^3q^2+384\alpha f_{1}r_{+}^4\Lambda q^2-2304\alpha^2 f_{1}^3r_{+}^4\Lambda -96\alpha f_{1}r_{+}^4-168\alpha f_{1}^2r_{+}^5-19r_{+}^8f_{1}-12r_{+}^9\nonumber\\
&\Lambda -20q^2r_{+}^5+6912\alpha^3 f_{1}^5-96\alpha f_{1}\Lambda r_{+}^6-96\alpha f_{1}q^2 r_{+}^2+768\alpha\Lambda f_{1}^2r_{+}^7+768\alpha f_{1}^2q^2r_{+}^3+192\alpha\nonumber\\
& f_{1}\Lambda^2 r_{+}^8+16r_{+}^7\Big),
\end{align}
and $f_{4}$ is provided in Appendix \ref{sec:Appendix}.
Notice that $f_1$ is not determined by the field equation itself. We now consider the thermodynamics of these black hole solutions.  For a static space-time, we have a timelike Killing vector
$ \xi=\partial_{t} $ everywhere outside the horizon. The black hole temperature can be determined by 
\begin{align}\label{eq20}
T &=\left. \dfrac{f^{'}(r)}{4\pi}\right\vert_{r_{+}}
= \dfrac{f_{1}}{4\pi}.
 \end{align}
Then, the entropy of the black hole is according to the Wald formula  \cite{Wald1, Wald2}
\begin{align}\label{eqqqentropy}
S=-2\pi\int_{r=r_+}d^{2}x\sqrt{\eta} \mathcal{P}^{a b c d}\epsilon_{a b}\epsilon_{c d}&=\dfrac{\mathcal{A}}{4}\left[1+6\alpha \left(\dfrac{f^{2}_{1}}{r_{+}^2}+\dfrac{4f_{1}}{r_{+}^3}\right)\right],
\end{align}
where $\mathcal{A}$ is the area of black hole and $\mathcal{P}_{a b c d}$ is the following
\begin{align}
\mathcal{P}_{a b c d}&=g_{a [c}\;g_{b]d}+6\alpha \Biggl[ \mathcal{R}_{a d}\mathcal{R}_{b c}-\mathcal{R}_{a c}\mathcal{R}_{b d}+g_{bd}\mathcal{R}_{a}^{e}\mathcal{R}_{c e}-g_{a d}\mathcal{R}_{b}^{e}\mathcal{R}_{ce}-g_{b c}\mathcal{R}_{a}^{e}\mathcal{R}_{de}+g_{a c}\mathcal{R}_{b}^{e}\mathcal{R}_{de} \nonumber\\
&- g_{bd}\mathcal{R}^{ef}\mathcal{R}_{aecf}+g_{bc}\mathcal{R}^{ef}\mathcal{R}_{aedf}+
g_{ad}\mathcal{R}^{ef}\mathcal{R}_{becf}-3\mathcal{R}_{a}{}^{e}{}_{d}{}^{f}\mathcal{R}_{becf}-g_{ac}\mathcal{R}^{ef}\mathcal{R}_{bedf} \nonumber \\ 
&+ 
3\mathcal{R}_{a}{}^{e}{}_{c}{}^{f}\mathcal{R}_{bedf} +  \dfrac{1}{2}\mathcal{R}_{ab}{}^{ef}\mathcal{R}_{cdef} \Biggr].
\end{align}
The chemical potential corresponding to the coupling constant $\alpha$ which has been obtained in Appendix \ref{appa}, is given as \cite{Hajian:2023bhq} 
\begin{equation}\label{psialpha2}
\psi_{\alpha} =-\dfrac{f_{1}^{2}(6+r_{+}f_{1})}{2r_{+}}.
\end{equation}
Since $f_{1}=4\pi T$, therefore $\psi_{\alpha}$ is always negative.
Other well-known thermodynamic quantities are the chemical potentials corresponding to the electric charge and the cosmological constant, which are given below \cite{Chernyavsky:2017xwm}
\begin{equation}
\psi_{q}=\dfrac{q}{r_{+}},\;\;\;\;\;\;\psi_{\Lambda}=\dfrac{r_{+}^3}{6}.
\end{equation}
For a black hole solution with thermodynamics parameters $M$, $q$, $\Lambda$ and $\alpha$ in Einstein Cubic Maxwell theory, the first law of thermodynamics and the Smarr formula can be written as 
\begin{equation}\label{eqfirstlaw}
dM=TdS+\psi_{\alpha} d\alpha -\psi_{\Lambda} d\Lambda +\psi_{q} dq,
\end{equation}
\begin{equation}\label{eq26}
M=2TS+4\alpha\psi_{\alpha} +2\psi_{\Lambda}\Lambda+q\psi_{q}.
\end{equation}
By substituting all the thermodynamic potentials into the Smarr formula, we obtain the following relation
\begin{equation}\label{eqmass}
M(r_{+},\alpha ,\Lambda ,q)=\dfrac{1}{2}f_{1}(r_{+},\alpha ,\Lambda ,q)r_{+}^{2}+\alpha f^{3}_{1}(r_{+},\alpha ,\Lambda ,q)+\dfrac{\Lambda r_{+}^3}{3}+\dfrac{q^2}{r_{+}},
\end{equation}
yielding the mass parameter as $M(r_{+},\alpha ,\Lambda,q)$.
 We now impose the first law \eqref{eqfirstlaw}, which becomes
\begin{align}\label{flaw1}
\dfrac{\partial M}{\partial r_{+}}dr_{+}+\dfrac{\partial M}{\partial \alpha}d\alpha + \dfrac{\partial M}{\partial \Lambda}d\Lambda + \dfrac{\partial M}{\partial q}dq &= T \dfrac{\partial S}{\partial r_{+}}dr_{+}+T \dfrac{\partial S}{\partial \alpha}d\alpha+T \dfrac{\partial S}{\partial \Lambda}d\Lambda \nonumber\\
&+ T \dfrac{\partial S}{\partial q}dq+\psi_{\alpha}   d\alpha -\psi_{\Lambda} d\Lambda +\psi_{q} dq,
\end{align}
yielding
\begin{align}\label{flawrpp}
&\dfrac{\partial M}{\partial r_{+}}-T\dfrac{\partial S}{\partial r_{+}}=0,\hspace{0.5cm}\Longrightarrow\nonumber\\
&\dfrac{1}{2}\dfrac{\partial f_{1}(r_{+},\alpha,\Lambda,q)}{\partial r_{+}}r_{+}^{2}+\dfrac{6\alpha f_{1}(r_{+},\alpha,\Lambda,q)}{r_{+}}\left(\dfrac{f_{1}(r_{+},\alpha,\Lambda,q)}{r_{+}}-\dfrac{\partial f_{1}(r_{+},\alpha,\Lambda,q)}{\partial r_{+}}\right)+\nonumber\\
&\dfrac{1}{2}f_{1}(r_{+},\alpha,\Lambda,q)r_{+}+\Lambda r_{+}^2-\dfrac{q^2}{r_{+}^2}=0,
\end{align}
and
\begin{align}\label{flawqq}
&\dfrac{\partial M}{\partial \alpha}-T\dfrac{\partial S}{\partial \alpha}-\psi_{\alpha}=0,\hspace{0.5cm}\Longrightarrow\nonumber\\
&\dfrac{1}{2}\dfrac{\partial f_{1}(r_{+},\alpha,\Lambda,q)}{\partial\alpha}r_{+}^{2}-\dfrac{3f^{2}_{1}(r_{+},\alpha,\Lambda,q)}{r_{+}}-\dfrac{6\alpha f_{1}(r_{+},\alpha,\Lambda,q)}{r_{+}}\dfrac{\partial f_{1}(r_{+},\alpha,\Lambda,q)}{\partial\alpha}=0,\\
&\dfrac{\partial M}{\partial \Lambda}-T\dfrac{\partial S}{\partial \Lambda}+\psi_{\Lambda}=0,\hspace{0.5cm}\Longrightarrow\nonumber\\
&\dfrac{1}{2}\dfrac{\partial f_{1}(r_{+},\alpha,\Lambda,q)}{\partial\Lambda}r_{+}^{2}+\dfrac{r_{+}^3}{2}-\dfrac{6\alpha f_{1}(r_{+},\alpha,\Lambda,q)}{r_{+}}\dfrac{\partial f_{1}(r_{+},\alpha,\Lambda,q)}{\partial\Lambda}=0,\\
&\dfrac{\partial M}{\partial q}-T\dfrac{\partial S}{\partial q}-\psi_{q}=0,\hspace{0.5cm}\Longrightarrow\nonumber\\
&\dfrac{1}{2}\dfrac{\partial f_{1}(r_{+},\alpha,\Lambda,q)}{\partial q}r_{+}^{2}+\dfrac{q}{r_{+}}-\dfrac{6\alpha f_{1}(r_{+},\alpha,\Lambda,q)}{r_{+}}\dfrac{\partial f_{1}(r_{+},\alpha,\Lambda,q)}{\partial q}=0,\label{flawqq}
\end{align}
as differential equations that must be satisfied by $f_{1}(r_{+},\alpha,\Lambda,q)$.
By solving differential equations (\ref{flawrpp})-\eqref{flawqq}, one can obtain two solutions for $f_{1}$ as follows


\begin{equation}\label{eq20q}
{ f^{\pm}_{1}(r_{+},\alpha, \Lambda, q)=\dfrac{r_{+}^3\pm\sqrt{r_{+}^{6}+24\alpha q^2+24\Lambda\alpha r_{+}^4+48\alpha c_{1}r_{+}^2}}{12\alpha},}
\end{equation}
here, $c_{1}$ is an integration constant. By expanding $f^{\pm}_{1}(r_{+},\alpha, \Lambda, q)$ around small $\alpha$ one can find only $f^{-}_{1}(r_{+},\alpha,\Lambda,q)$ has a smooth $\alpha\to 0$ limit. Now, we can put $f_{1}^{-}$ in the thermodynamic quantities. At first by inserting \eqref{eq20q} into the temperature, one can get
\begin{equation}\label{eqtemm}
T=\dfrac{r_{+}^3-\sqrt{r_{+}^{6}+24\alpha q^2+24\Lambda\alpha r_{+}^4+48\alpha c_{1}r_{+}^2}}{48\pi \alpha}.
\end{equation}
In a limit $\alpha\ll 1$, we have 
\begin{equation}\label{eqtemm1}
T\approx -\dfrac{c_{1}}{2\pi r_{+}}-\dfrac{\Lambda r_{+}}{4\pi}-\dfrac{q^2}{4\pi r_{+}^3}+\mathcal{O}(\alpha).
\end{equation}
To obtain the standard Schwarzschild temperature at the leading order, we set $c_{1}=-1/2$. One gets extremal conditions by setting the temperature to zero. This gives
\begin{align}\label{eqexttem}
q_{ext} &= \sqrt{1-\Lambda r_{+}^2}r_{+},
\end{align}
which is independent of $\alpha$.
For $\Lambda=0$, the temperature \eqref{eqtemm} becomes
\begin{equation}
T=\dfrac{r_{+}^3-\sqrt{r_{+}^{6}+24\alpha q^2-24\alpha r_{+}^2}}{48\pi \alpha}.
\end{equation}
This is the same as the temperature obtained in \cite{Frassino:2020zuv}.
In the case of $\Lambda=0$ and $q=0$, the temperature \eqref{eqtemm} will be
\begin{equation}
T=\dfrac{r_{+}(r_{+}^2-\sqrt{r_{+}^{4}+24\alpha c_{1}})}{48\pi \alpha},
\end{equation}
which corresponds to the temperature in \cite{Bueno:2016lrh,Hennigar:2018hza,Hennigar:2016gkm}. Inserting \eqref{eq20q} into \eqref{eqmass}, one can get the ADM mass of the black hole as
\begin{equation}\label{eqmass0}
M=\dfrac{r_{+}^{10}+18\Lambda\alpha r_{+}^{8}+432\alpha^2 q^2+18\alpha(8\alpha\Lambda +q^2)r_{+}^4-(r_{+}^7+12r_{+}^3\alpha +6\alpha r_{+}q^2+6\alpha\Lambda r_{+}^5)\sqrt{\mathfrak{C}}}{432r_{+}\alpha^2},
\end{equation}
here
\begin{equation}
\mathfrak{C}=r_{+}^6+24\alpha q^2+24\Lambda\alpha r_{+}^4-24\alpha r_{+}^2.
\end{equation}
In the small $\alpha$ limit, the mass becomes
\begin{align}\label{eqmass1}
M\approx \dfrac{r_{+}}{2}+\dfrac{q^2}{2r_{+}}-\dfrac{\Lambda r_{+}^3}{6}+\Big(\dfrac{6q^4}{r_{+}^7}-\dfrac{q^6}{r_{+}^9}-\dfrac{3q^4\Lambda}{r_{+}^5}+&\dfrac{12q^2\Lambda}{r_{+}^3}-\dfrac{3q^2\Lambda^2}{r_{+}}-\dfrac{9q^2}{r_{+}^5}+6\Lambda^2 r_{+}\nonumber\\
&-\Lambda^3 r_{+}^3-\dfrac{9\Lambda}{r_{+}}+\dfrac{4}{r_{+}^3}\Big)\alpha+\mathcal{O}(\alpha^2).
\end{align}
The first term in the above expression is due to Einstein's gravity, and the second one is the correction term from the cubic gravity.
By inserting \eqref{eqexttem} into the \eqref{eqmass0}, the extremal mass and electric charge are given as 
\begin{equation}
M_{ext}=r_{+}-\dfrac{2}{3}\Lambda r_{+}^3,\;\;\;q_{ext}=\sqrt{1-\Lambda r_{+}^2}r_{+},
\end{equation}
in which for $\Lambda=0$, one can get $M_{ext}=q_{ext}=r_{+}$. This is an extremal point for a standard Reissner-Nordstr\"om black hole. This means that the extremal point of the black hole is the same for Einstein's gravity and Einstein's cubic gravity.
In the case of $q=\Lambda=0$, one can get
\begin{equation}
M=\dfrac{r_{+}^{3}\left(r_{+}^2-\sqrt{r_{+}^{4}-24\alpha}\right)\left(r_{+}^{4}+24\alpha-r_{+}^{2}\sqrt{r_{+}^{4}-24\alpha}\right)}{864\alpha^2},
\end{equation}
which is consistence with the mass in \cite{Bueno:2016lrh,Hennigar:2018hza,Hennigar:2016gkm}.
Inserting \eqref{eq20q} into the \eqref{eqqqentropy}, the black hole entropy reads
\begin{equation}
S=\dfrac{\pi(r_{+}^7+12\alpha\Lambda r_{+}^5+24\alpha r_{+}^3+12\alpha q^2r_{+}-(r_{+}^4+24\alpha)\sqrt{\mathfrak{C}})}{12\alpha r_{+}}.
\end{equation}
In the small $\alpha$, it is given as
\begin{equation}
S\approx \pi r_{+}^{2}+\Big(\dfrac{6\pi q^4}{r_{+}^6}+\dfrac{12\pi q^2\Lambda}{r_{+}^2}-\dfrac{36\pi q^2}{r_{+}^4}+6\pi r_{+}^2\Lambda^2-36\pi\Lambda+\dfrac{30\pi}{r_{+}^2}\Big)\alpha +\mathcal{O}(\alpha^2).
\end{equation}
The first term is the black hole entropy in Einstein's gravity and the second term is a contribution from the cubic gravity. In the case of $\Lambda = 0$, this entropy reduces to those obtained previously in \cite{Frassino:2020zuv}.
For $q=\Lambda=0$, the entropy of the black hole becomes
\begin{equation}
S=\dfrac{\pi(r_{+}^{4}+24\alpha)(r_{+}^{2}-\sqrt{r_{+}^{4}-24\alpha})}{12\alpha},
\end{equation}
which agrees as the results previously obtained in \cite{Bueno:2016lrh,Hennigar:2018hza,Hennigar:2016gkm}. This shows that by having thermodynamic potentials, without having an explicit form of the metric function, and without the need to solve the field equations, it is possible to study the thermodynamics of the black hole. In this way, one can study the thermodynamics of every theory that admits the solutions with one metric function, including Einstein-quartic gravity presented in \cite{Ahmed:2017jod, Rezzolla:2014mua}. 

With $f_{1}$ in hand from \eqref{eq20q}, now the full metric can be determined using a continued fraction expansion \cite{Wilson-Gerow:2015esa, Kokkotas:2017zwt, Konoplya:2019ppy, Zinhailo:2018ska}. We write 
\begin{equation}\label{eq17}
f(r)=xA(x),\hspace{1cm} x= 1- \frac{r_+}{r},
\end{equation}
with
\begin{align}
A(x) &=1-\epsilon(1-x)+(a_{0}-\epsilon)(1-x)^{2}+\dfrac{a_{1}(1-x)^{3}}{1+\dfrac{a_{2}x}{1+\dfrac{a_{3}x}{1+\dfrac{a_{4}x}{1+...}}}}.
\label{Ax}
\end{align}
In this work, we truncate the continued fraction by an order of four.
By expanding (\ref{eq17}) near the horizon ($ x\to 0 $) and
the asymptotic  region ($ x\to 1 $)  we obtain
\be
\epsilon=\dfrac{2M}{r_{+}}-1, \qquad a_{0}=\dfrac{q^2}{r_{+}^2},\qquad a_{1}=-1-a_{0}+2\epsilon+r_{+}f_{1}
\ee
for the lowest-order expansion coefficients, with the remaining
$a_i$ given in terms of $(r_+, f_1)$; we provide these expressions in the Appendix \ref{sec:Appendix}. 
In Fig. \ref{frmetr1}, we display the metric function $f(r)$ using \eqref{eq17}, \eqref{eq20q}, \eqref{eqmass0}, and employing up to $a_{4}$ term in continued fraction expansion. We also show the differences between the numerical solution and the continued fraction approximation. Despite some disagreement at small $r$, it is clear that the approximate analytic solution and the numeric one agree very well after a certain value of $r$. We also display an approximate solution at large $r$ using \eqref{eqasymp} up to order $r^{-100}$ (dashed red curve), which agrees very well at large $r$ as expected. 
\begin{figure}[H]\hspace{0.4cm}
\centering
\subfigure{\includegraphics[width=0.6\columnwidth]{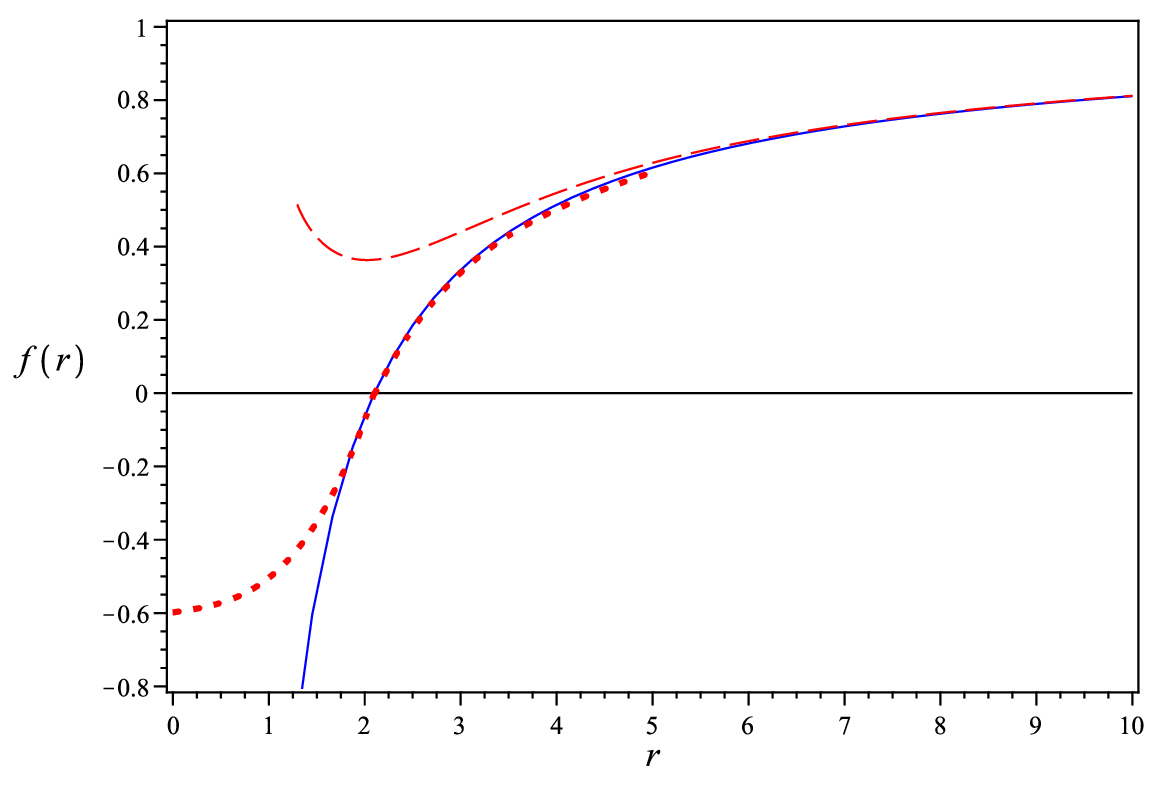}}
\caption{Comparison of numeric solution (\textcolor{red}{dotted line}) and continued fraction approximation (\textcolor{blue}{solid line}) \eqref{eq17} for $\alpha=-0.3,r_{+}=2,\Lambda=q=0$. In the continued fraction, we keep expansion terms up to $a_{4}$. The \textcolor{red}{dashed red} curve is the asymptotic solution \eqref{eqasymp}, including terms up to order $r^{-100}$. 
} 
\label{frmetr1}
\end{figure}
It should be noted that, in this figure and the following figures, we have used \eqref{eqmass0} to obtain the mass.
In the following, we investigate the superradiance of a charged scalar field
around a static spherically symmetric black hole solution by using the outcomes of this section. 

\section{{Superradiance of charged scalar field}}\label{sec4}

In a curved spacetime, a massless scalar field $\phi(t,r,\theta,\phi)$ minimally coupled to vector potential $A_{\mu}$ is described by the Klein-Gordon equation \cite{Brito:2015oca}

\begin{equation}
(\nabla_{\mu}-iQA_{\mu})(\nabla^{\mu}-iQA^{\mu}) \Phi=0,
\end{equation}
where $Q$ is the charge of the scalar field. Here, we assume an electrical gauge potential $A_{\mu}=\left(-\frac{q}{r},0,0,0\right)$. In the static black hole spacetime described by \eqref{metform}, we make the following field decomposition 
\begin{equation}
\Phi(t,r,\theta,\phi)=e^{-i\omega t}R(r)Y(\theta,\phi),
\end{equation}
where $Y(\theta,\phi)$ is spherical harmonics. The ordinary differential equations that govern $R(r)$ is as follows,
\begin{equation}\label{eqq9R}
\dfrac{d^{2}R(r)}{dr^2}+\left(\dfrac{f^{\prime}}{f}+\dfrac{2}{r}\right)\dfrac{dR(r)}{dr}+\left(\dfrac{\omega^2}{f^2}-\dfrac{l(l+1)}{r^2f}-\dfrac{2\omega qQ}{rf^2}+\dfrac{q^2Q^2}{r^2f^2}\right)R(r)=0,
\end{equation}
where $\omega$ is the frequency of the charged scalar field, $q$ is the charge of BH, and $l=0,1,2,...$. 

\begin{figure}[H]\hspace{0.4cm}
\centering
\subfigure{\includegraphics[width=0.6\columnwidth]{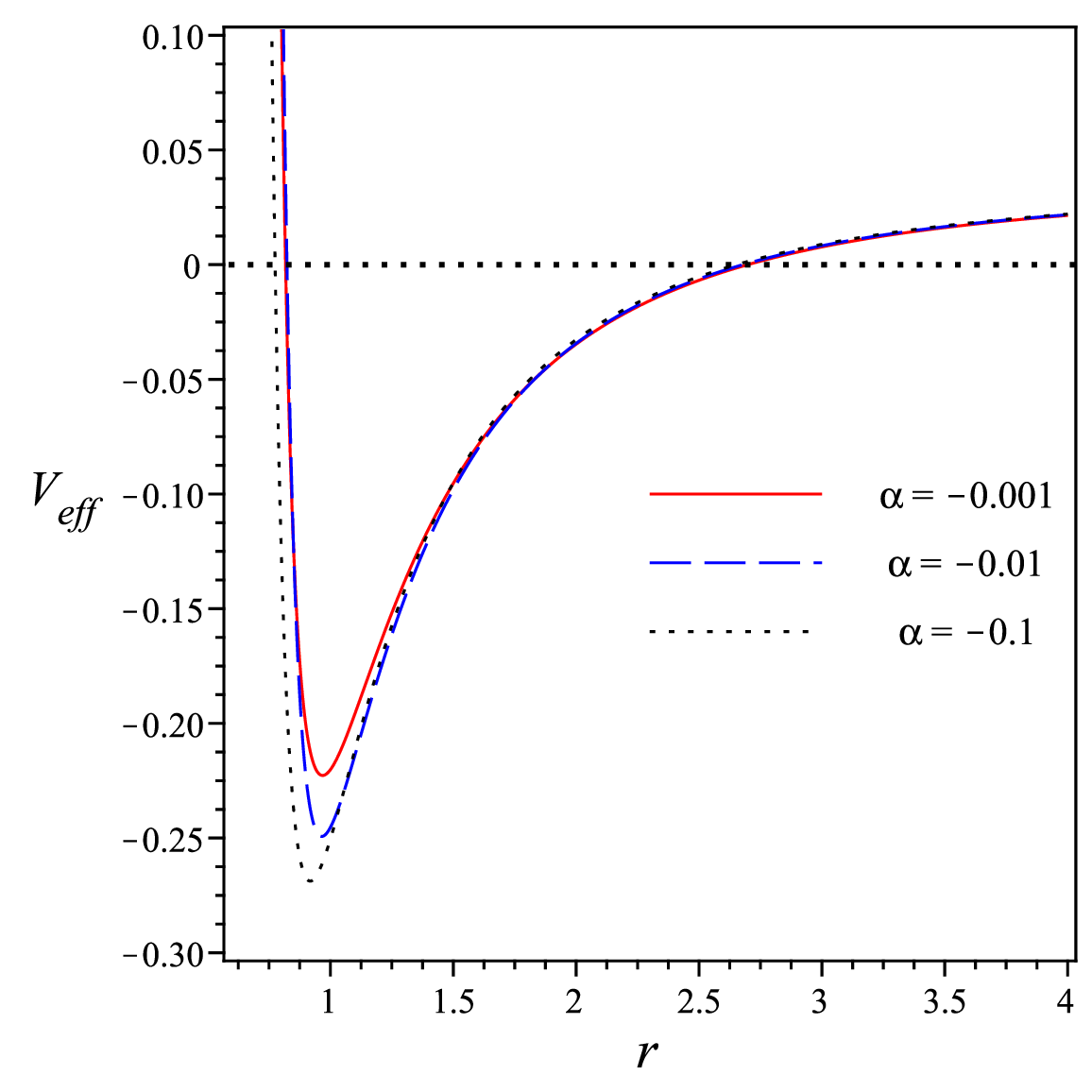}}
\caption{The function $V_{eff}(r)$ \eqref{eqeffect} of charged scalar field in the background of static BH, in terms of $r$ for $q=0.2, Q=0.4,\omega =0.2,l=0,r_{+}=0.3, \Lambda=0$. 
} 
\label{frmetr11}
\end{figure}

In terms of the tortoise coordinate $r_{\star}$, the radial function satisfies the Schr\"odinger-like equation 
\begin{equation}\label{eqqshrod}
\dfrac{d^2u(r)}{dr_{\star}^{2}}+V_{eff}(r)u(r)=0,
\end{equation}
where
\begin{equation}
r_{\star}=\int \dfrac{dr}{f},\;\;\;\;\;u(r)=rR(r),
\end{equation}
and the effective potential $V_{eff}$ which encodes the curvature of the background and the properties of the test field is given as
\begin{equation}\label{eqeffect}
V_{eff}(r)=\omega^2 -\left[\dfrac{ff^{\prime}}{r}+\dfrac{l(l+1)f}{r^2}+\dfrac{2\omega qQ}{r}-\dfrac{q^2Q^2}{r^2}\right].
\end{equation}
From \eqref{eqqshrod}, in the region where $V_{eff}>0$, the function $u(r)$ is oscillatory, and in the region $V_{eff}<0$, it is exponential that is a growing or damping wave (superradiance happens in this region.). As the angular momentum $l$ increases, the height of the potential barrier decreases. Also, the peak of the effective potential decreases (increases) with $q>0$ and ($q<0$).   This is anticipated because particles with the same charge are repelled and less absorbed by the BH. The effective potential does not depend explicitly on the coupling constant of the theory. The coupling constant affects the effective potential through the metric function instead. 
In figure \ref{frmetr11}, the behavior of the effective potential \eqref{eqeffect} is shown. As can be seen, by increasing $\alpha$, the height of the effective potential increases.

\begin{figure}[H]\hspace{0.4cm}
\centering
\subfigure{\includegraphics[width=0.6\columnwidth]{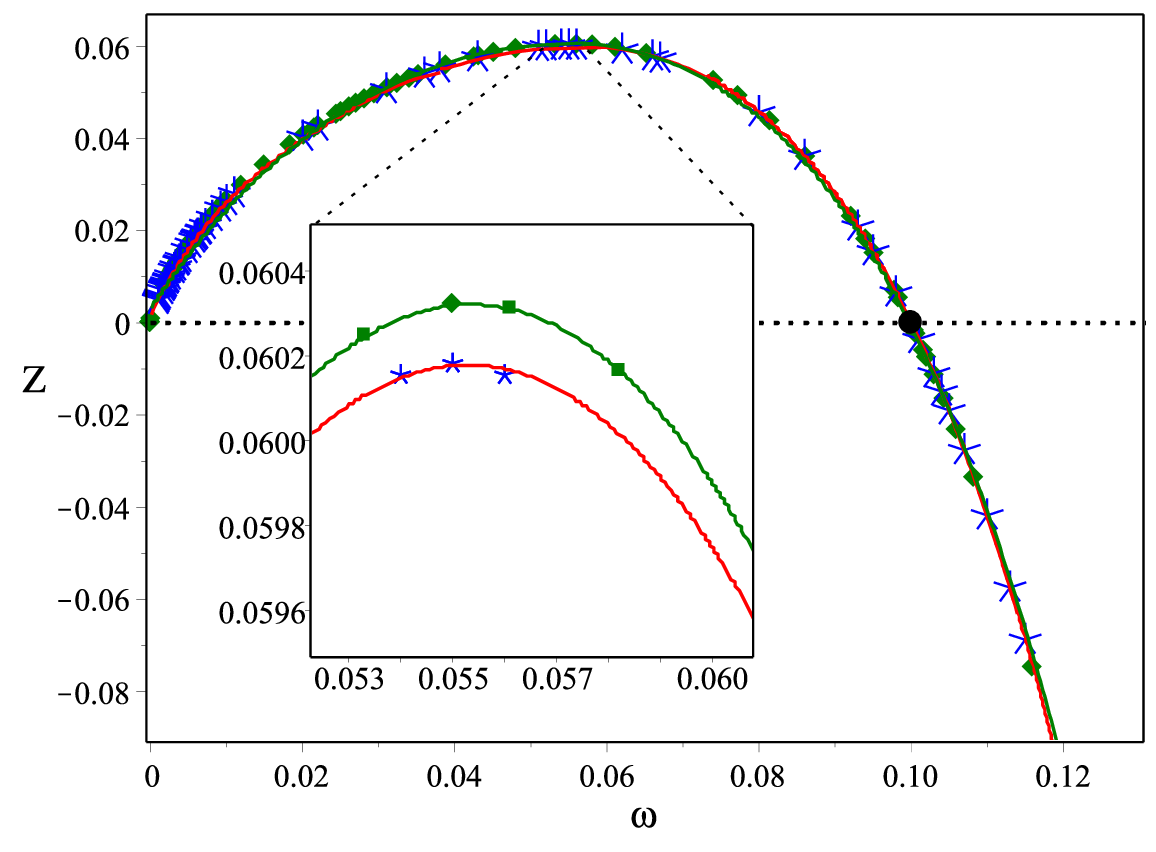}}
\caption{Superradiance of a complex massless scalar field with $l=0$ and $\alpha=\textcolor{green}{-0.05},\textcolor{red}{-0.1},q=0.5,Q=0.6,r_{+}=3,\Lambda=0$. The red solid line is the curve fitted with the numerical results.
} 
\label{wpert00}
\end{figure}

Now, we consider the asymptotic behavior of the effective potential and solution. For $r\to r_{+}$, the effective potential becomes $V_{eff}\to k^{2}_{+}=(\omega-qQ/r_{+})^2$, and the solution for the equation \eqref{eqqshrod} becomes
\begin{equation}\label{eqqu70}
u_{+}(r_{\star})=\mathcal{A}_{t}e^{-ik_{+}r_{\star}}.
\end{equation}
At the horizon, only an ingoing wave is allowed. For $r\to\infty$, the effective potential becomes $V_{eff}=k_{\infty}^{2}=\omega^2$, and the asymptotic solution is given as
\begin{equation}\label{eqqu72}
u_{\infty}(r_{\star})=\mathcal{A}_{i}e^{-ik_{\infty}r_{\star}}+\mathcal{A}_{r}e^{ik_{\infty}r_{\star}}.
\end{equation}
The boundary condition \eqref{eqqu72} represents an incoming wave with the amplitude $\mathcal{A}_{i}$ coming from spatial infinity. The incoming wave then scatters off the effective potential, which gives rise to a reflected and transmitted wave \eqref{eqqu70} with the amplitudes $\mathcal{A}_{r}$ and $\mathcal{A}_{t}$ respectively.
The Wronskian for regions near the event horizon is obtained as follows
\begin{equation}
W_{+}=u_{+}\dfrac{du^{\star}_{+}}{dr_{\star}}-u^{\star}_{+}\dfrac{du_{+}}{dr_{\star}}=2ik_{+}\vert \mathcal{A}_{t} \vert^{2},
\end{equation}
and the Wronskian for regions near infinity is obtained as follows 
\begin{equation}
W_{\infty}=u_{\infty}\dfrac{du^{\star}_{\infty}}{dr_{\star}}-u^{\star}_{\infty}\dfrac{du_{\infty}}{dr_{\star}}=ik_{\infty}\left(\vert \mathcal{A}_{i}\vert^2-\vert \mathcal{A}_{r}\vert^2\right),
\end{equation}
where $u^{\star}$ is the complex conjugate of $u$. Now, by equating the Wronskian for regions near the event horizon with its other counterparts at infinity, we arrive at the following conditions
\begin{equation}
\vert \mathcal{A}_{i}\vert^2-\vert \mathcal{A}_{r}\vert^2 =\dfrac{2k_{+}}{k_{\infty}}\vert \mathcal{A}_{t} \vert^{2}.
\end{equation}
In order for the superradiance to take place, the amplitude of the reflected wave must exceed the amplitude of the incident wave i.e., $\vert\mathcal{A}_{i}\vert^2-\vert \mathcal{A}_{r}\vert^2<0,$ so the following frequency conditions must be met
\begin{equation}\label{condsuper}
0 <\omega <\dfrac{qQ}{r_{+}}.
\end{equation}
The critical frequency $\omega_{c} \equiv qQ/r_{+}$ is the same as those of Reissner-Nordstr\"om (RN) BH ($\omega_{c}=\omega_{c}^{RN}$) \cite{Crispino:2009ki}. In terms of wavelength $\lambda$ of the field, it can be rewritten as follows:
\begin{equation}
   \dfrac{2\pi}{qQ}\leq\dfrac{\lambda}{r_{+}}.
\end{equation}
In our cases since $qQ<1$, therefore $\lambda\gg r_{+}$. For $Q>2\pi$ and $0<q<1$, we have $\lambda\sim r_{+}$.
To quantify the superradiance, one can define an amplification factor as follows
\begin{equation}
Z_{\omega l}=\dfrac{\vert \mathcal{A}_{r}\vert^2}{\vert \mathcal{A}_{i}\vert^2}-1. \label{ampfac}
\end{equation}
This measures the fractional gain or loss of energy in a scattered wave, with positive values of $Z_{\omega l}$ corresponding to superradiant amplification.
The amplification factor $Z_{\omega l}$ with a given $\omega$ can be computed by numerically integrating equations \eqref{eqqshrod}-\eqref{eqeffect}. With the given metric function \eqref{eq17}, the tortoise coordinate can then be evaluated. Thus, the wave equation \eqref{eqqshrod} can be directly integrated, and the amplification factor can be computed via \eqref{ampfac}. In Fig. \ref{wpert00}, we display the amplification factor $Z_{\omega l}$ as a function of the frequency $\omega$ for $l = 0$ and $\alpha=\textcolor{green}{-0.05},\textcolor{red}{-0.1}$. The black dot denotes the superradiance threshold, $Z_{\omega l}= 0$, which is found to be consistent with $\omega= \omega_{c}=0.1$. When $\omega<\omega_{c}$, superradiance occurs since $Z_{\omega l}>0$. By decreasing $\alpha$, the amplification factor decreases. 
 
\begin{figure}[H]\hspace{0.4cm}
\centering
\subfigure{\includegraphics[width=0.45\columnwidth]{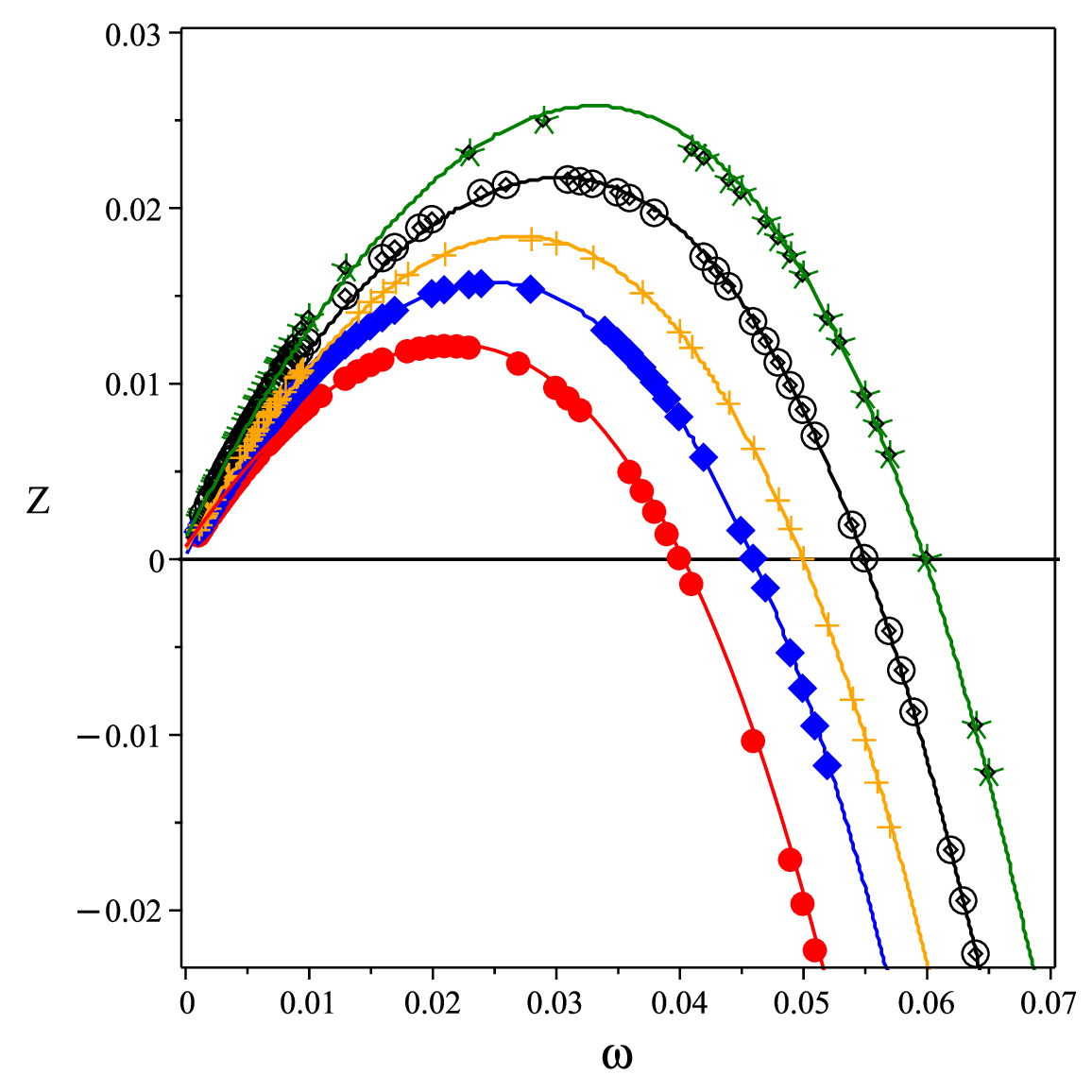}}
\subfigure{\includegraphics[width=0.45\columnwidth]{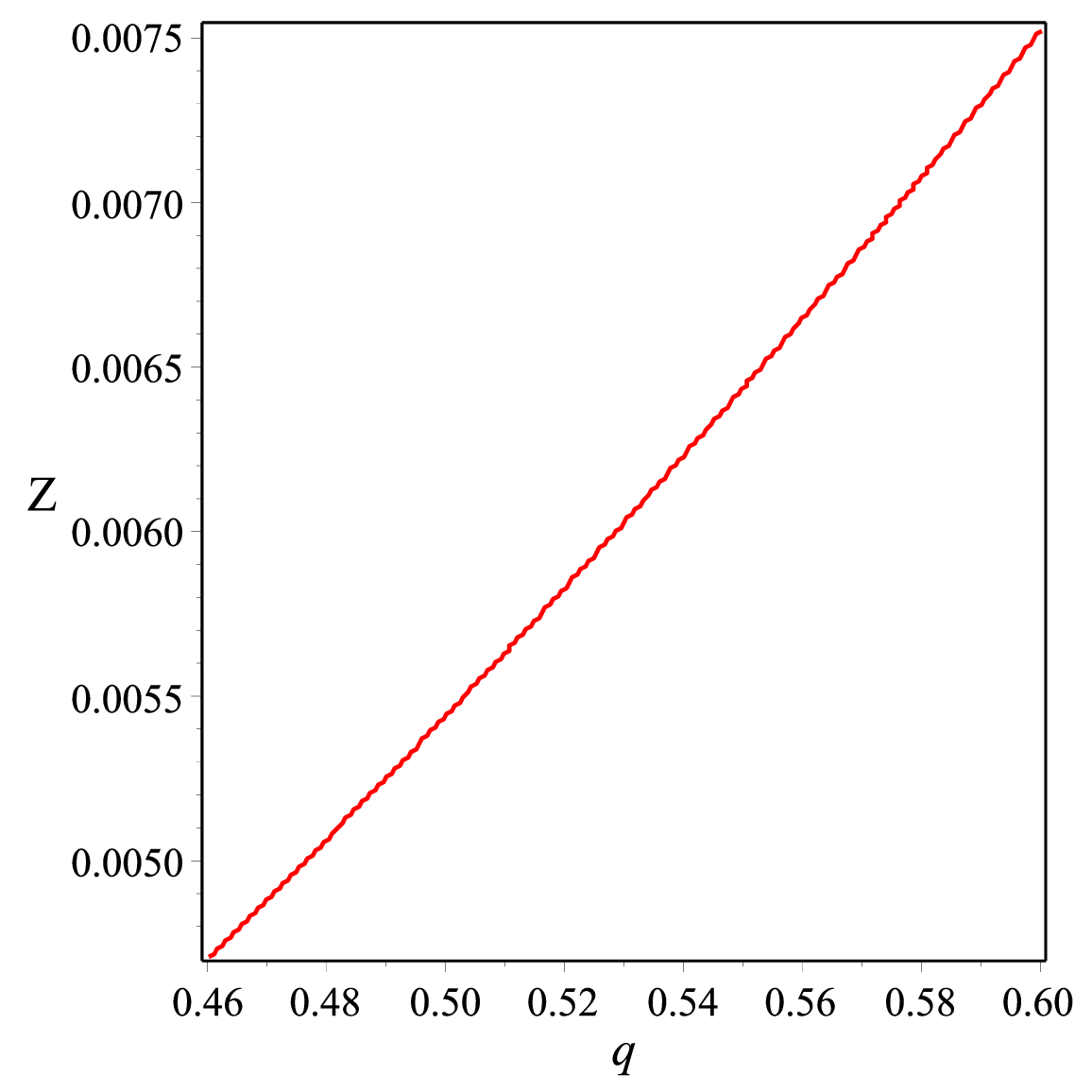}}
\caption{Superradiance ($Z_{\omega l}$) of a complex massless scalar field in terms of $\omega$ with $l=0,r_{+}=3,\alpha=-0.1,Q=0.3,\Lambda=0$ and $q=\tcr{0.46},\tcb{0.5},\textcolor{orange}{0.55},0.6,\textcolor{green}{0.65}$ (left) and $Z_{\omega l}$ vs $q$ for $r_{+}=3,\Lambda=0,\alpha=-0.1,\omega=1.2\times 10^{-3}$ (right). The colored solid lines are the curves fitted with the numerical results.} 
\label{wpert1}
\end{figure}

In the left panel of Fig. \ref{wpert1}, the behavior of the amplification factor in terms of $\omega$ for different values of $q$ has been shown. As can be seen, the peak value of superradiance magnification increases with $q$. This shows that when $q$ increases, the energy extraction efficiency of complex massless scalar particles in the charged black hole increases, and the vanishing amplification factor means that the black hole will reflect all the incoming particles back, and there is no transfer of mass and charge between the black hole and scalar field. 
In the right panel, we particularly emphasize the fact that the amplification factor increases with $q$. This is due to the presence of a repulsive electromagnetic interaction for $qQ>0$ competing with the gravitational interaction, causing the decrease of the absorption. If the charges of the field $Q$ and the BH $q$ have opposite signs, the amplification is smaller than for the chargeless field case, due to the Lorentz attraction between the charges \cite{Crispino:2009ki},\cite{Hod:2012wmy}.

\begin{figure}[H]\hspace{0.4cm}
\centering
\subfigure{\includegraphics[width=0.65\columnwidth]{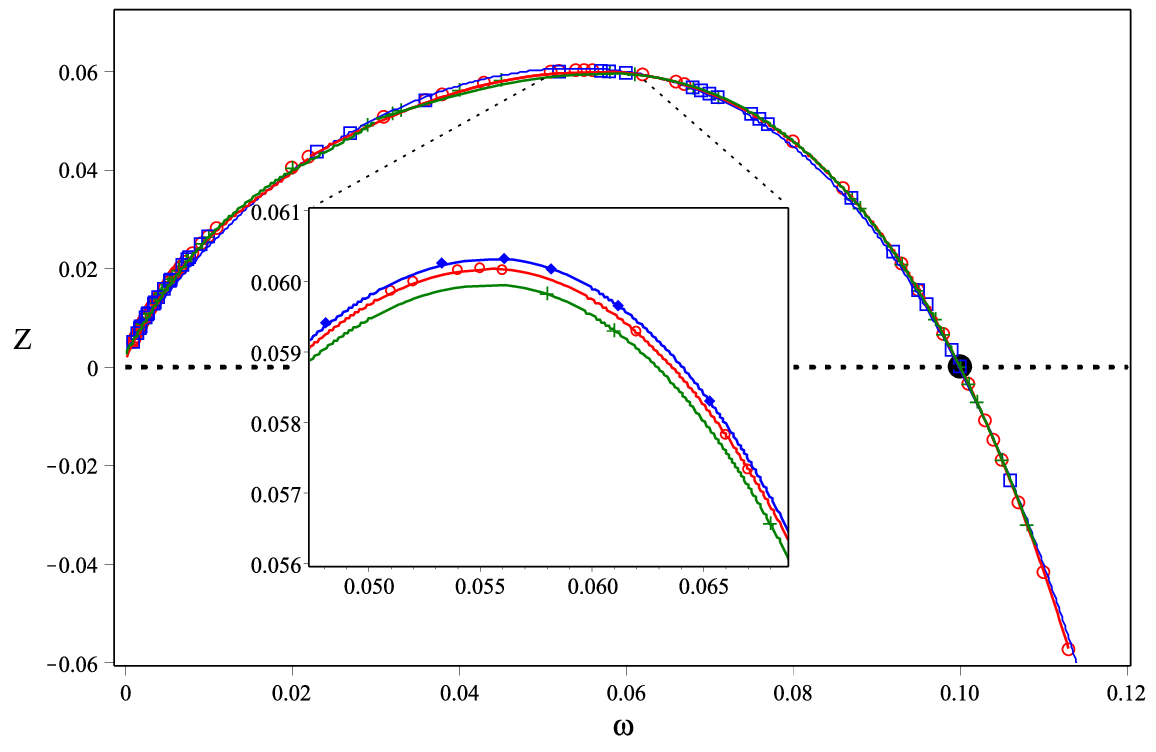}}
\caption{Superradiance ($Z_{\omega l}$) of a complex massless scalar field in terms of $\omega$ with $l=0,r_{+}=3,Q=0.6,q=0.5,\Lambda=0$ and $\alpha=\tcb{-0.1},\tcr{-0.2},\tcg{-0.3}$ (left) and $Z_{\omega l}$ vs $\alpha$ for $\omega=1.2\times 10^{-3}$ (right). } 
\label{wpert11}
\end{figure}

In Fig. \ref{wpert11}, the behavior of the amplification factor in terms of $\omega$ for different values of $\alpha$ is shown. As can be seen, the amplification factor is affected by the coupling constant $\alpha$. Increasing $\alpha$, the amplification factor increases. One can interpret this effect in the following way. Since our theory of gravity is higher curvature gravity, therefore we are dealing with strong gravity. In this high-energy regime, on the potential barrier near the event horizon of the black hole, based on the uncertainty principle, in a short time, the rate of spontaneous particle-antiparticle creation can increase. Also, the incident beam stimulates pair creation at the barrier, which emits particles and antiparticles. Finally, all the particles incident with energy $\omega$ are reflected back with energy $\omega$, but because of pair creation, more particles with charge $Q$ and energy $\omega$ join the beam. Since more particles are reflected than incident, so superradiance has happened. So, if we interpret the superradiance scattering as extracted energy from the spacetime background, it should occur more in higher derivative gravity. Accordingly, and as a result, the coupling constant should increase the superradiance.
 

\begin{figure}[H]\hspace{0.4cm}
\centering
\subfigure{\includegraphics[width=0.4\columnwidth]{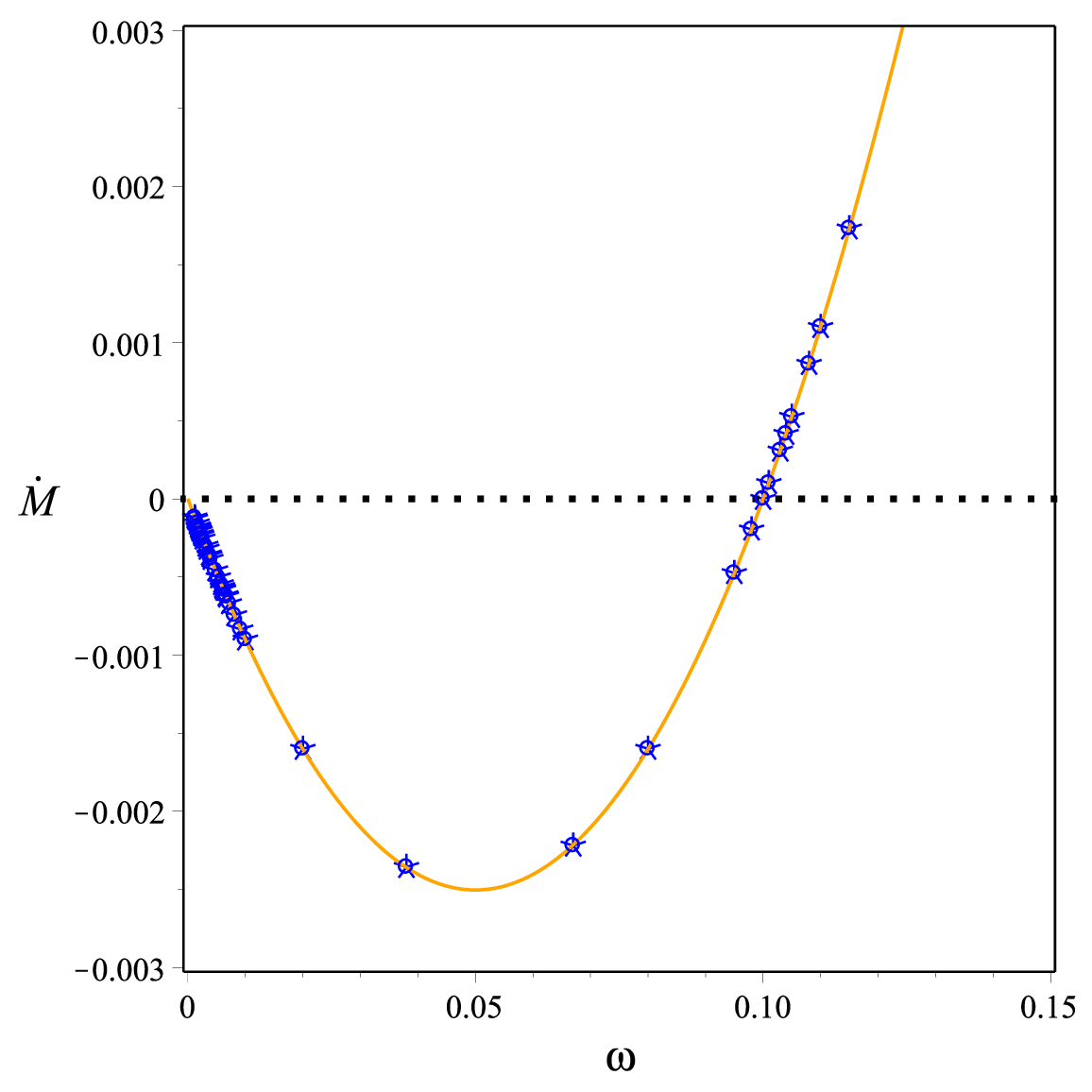}}
\subfigure{\includegraphics[width=0.4\columnwidth]{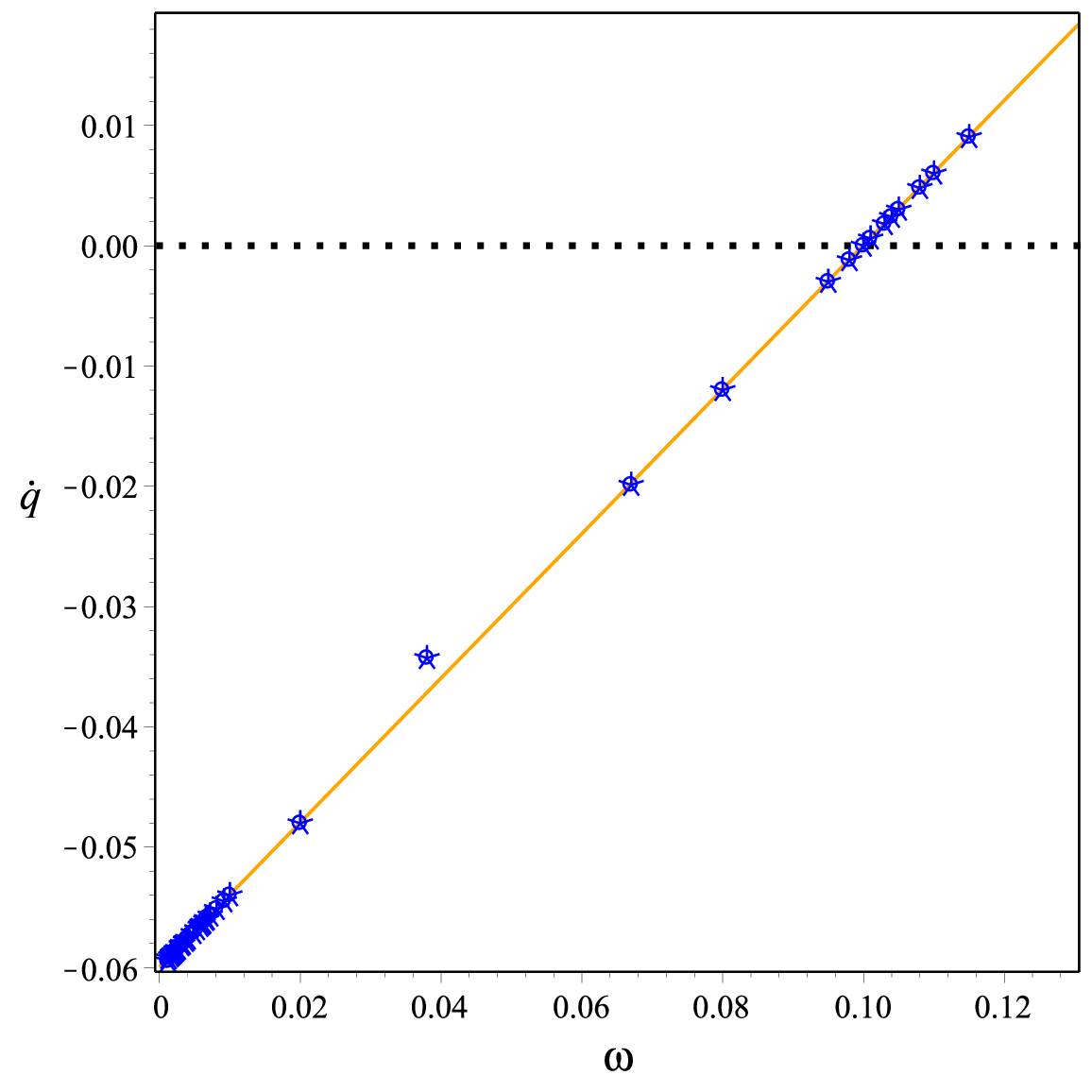}}
\caption{The rate of BH mass and charge \eqref{eqmdot} for the superradiant scattering of a scalar field around a BH in terms of $\omega$ for $r_{+}=3, \alpha=-0.1, Q=0.6,\Lambda=0$. The solid line is the curve fitted with the numerical results.
} 
\label{mqdot}
\end{figure}

 By considering the back-reaction (considering the second order of perturbation) of the scalar field on the spacetime and vector field, for a monochromatic wave at a large distance, one gets \cite{Brito:2015oca}
 \begin{equation}\label{eqmdot}
 \dot{M}=-\omega^2 \left(\vert \mathcal{A}_{r}\vert^2-\vert \mathcal{A}_{i}\vert^2\right),\;\;\;\;\dot{q}=-Q\omega (\vert \mathcal{A}_{r}\vert^{2}-\vert \mathcal{A}_{i}\vert^{2}).
 \end{equation}
Here, the dot is derivative with respect to $t$. As seen in Fig. \ref{mqdot}, in the superradiant regime, i.e, $\omega<\omega_{c}$ and $\vert \mathcal{A}_{r}\vert^{2}-\vert \mathcal{A}_{i}\vert^{2}>0$, both $\dot{M}$ and $\dot{q}$ have negative values. This means that the mass and charge of the black hole decrease, and energy comes out of the black hole and goes towards the exterior region. As $\omega\to\omega_{c}$, the $\dot{M}, \dot{q} \to 0$ and energy flux approaches zero. In contrast, beyond the superradiance regime, the mass and charge of the black hole increase (since $\dot{M},\dot{q}>0$). Even though superradiance seems to be the consequence of relativistic field equations in curved spacetime, it can also be understood by using the first and second laws of thermodynamics of black holes \cite{Bekenstein:1973mi}. Consider a charged black hole in equilibrium with entropy $S$, mass $M$, temperature $T$, electric charge $q$, and coupling constant $\alpha$. Suppose also that a wave packet with charge $Q$ and frequency $(\omega,\omega+d\omega)$ and azimuthal number $m$ is incident upon this BH. Assuming $\Lambda=0$, from the first law \eqref{eqfirstlaw}, the change in energy is


\begin{equation}\label{eqflwla}
dM=TdS+\psi_{\alpha} d\alpha +\psi_{q} dq.
\end{equation}
On the other hand, the interaction between a static charged BH and a wave with charge $Q$ leads to a change in the BH charge as
\begin{equation}
\dfrac{d q}{d M}=\dfrac{Q}{\omega}.
\end{equation}
Therefore, the \eqref{eqflwla} reads 
\begin{equation}\label{eqflwlmo}
dM=\dfrac{\omega T dS}{\omega- Q\psi_{q}}+\dfrac{\omega \psi_{\alpha}d\alpha}{\omega- Q\psi_{q}}.
\end{equation}
If $d\alpha=0$ and following the second law of thermodynamics $dS\geq0$, superradiance region ($\omega<Q\psi_q$) we have $dM<0$ i.e., the BH loses its mass. In the case where $d\alpha\neq0$ since $\psi_\alpha \leq 0$ from \eqref{psialpha2}, then the overall sign of $dM$ solely depends on the $d\alpha$. For $d\alpha<0$, we get $dM<0$ in the superradiance region. If $d\alpha >0$, then the second term in \eqref{eqflwlmo} is positive. Hence, for this case, $dM$ will be negative if 


\begin{equation}
\dfrac{dS}{d\alpha}\geq\dfrac{-\psi_{\alpha}}{T}\;\;\;\;\;\to\;\;\;\;\;r_{+}\geq q.
\end{equation}
Therefore, in the superradiance regime, the BH loses its mass and its charges as a consequence of the second law of thermodynamics and the effect of the coupling constant of the theory.

\section{Conclusion}\label{con}

In this work, we obtain the analytical approximate black hole solution of Einstein-cubic gravity using the continued fraction expansion. The continued fraction expansion accurately gives approximate black hole solutions that are regular everywhere outside the event horizon. To get these approximate solutions, we construct the metric function near the event horizon and at the asymptotic region. Then, the solutions are matched using continued-fraction expansion. In order to match the solutions from both ends, two unknown expansion parameters are determined by the first law of thermodynamics and the Smarr relation. With the obtained metric function, we investigate the superradiance of charged scalar perturbations, using direct numerical integration. We find that a charged scalar wave with frequency $\omega$ can be amplified superradiantly if $0<\omega <qQ/r_{+}$. We show that by increasing the $qQ>0$, the amplification factor increases, and this is the consequence of a repulsive electromagnetic field. Also, we have shown that by increasing the coupling constant of the theory, superradiance amplification slightly increases. We investigate how the black hole's mass and charge change with time by considering the second order of perturbations. We show that in the superradiance regime \eqref{condsuper} the mass and charge of the BH decrease, and this is in agreement with the non-zero amplification factor. Despite the fact that the coupling constant $\alpha$ does not appear explicitly in the effective potential, $\alpha$ affects the effective potential through the presence of the metric function $f$. This implicit dependency of $\alpha$ prevents the effective potential from having a double peak structure outside the black hole's horizon. Therefore, superradiance modes are not bounded and hence cannot develop superradiant instability. 

 
For future investigations, one can consider black holes' superradiance and superradiance instability using different probes in the presence of cosmological constants. The astrophysical BH are known to be rotating. Future studies of the superradiance of spinning BHs of this theory would clarify the difference between a charged static BH and a rotational one. 

\section*{Acknowledgements}
This research has received funding support from the NSRF via the Program Management Unit for Human Resource and Institutional Development, Research and Innovation grant number $B13F670063$.

\section*{Data Availability Statement}
No datasets were generated or analysed during the current study.

\appendix

\section{Chemical potential corresponds to coupling constant of theory}\label{appa}
In this appendix, we follow the approach in \cite{Hajian:2023bhq} and consider the first law of thermodynamics. Here, the coupling constant $\alpha$ is promoted to the scalar function $\alpha(x)$. The field strength $F_{\alpha}(x)$ is implemented in the following Lagrangian \cite{Hajian:2023bhq}
\begin{equation}
L=\dfrac{1}{16\pi}\left(R+\alpha(x)(P-F_{\alpha}(x))\right).
\end{equation}
By variation of the Lagrangian with respect to the field $\alpha$, an equation of motion is derived which implies the following on-shell relation
\begin{equation}
F_{\alpha}(x)=P.
\end{equation}
Therefore, the field strength can be written explicitly as
\begin{align}
{F}_{\alpha}(x)=-\dfrac{6f^{\prime}\sin(\theta)}{r^2}\left(rf^{\prime\prime}(4-4f+rf^{\prime})-2f^{\prime}(1+rf^{\prime}-f)\right) dt\wedge dr\wedge d\theta \wedge d\phi.
\end{align}
The gauge field whose field strength is calculated above is
\begin{align}
A_{\alpha}(x)=-\left[\dfrac{2f^{\prime 2}}{r}\left(6-6f+rf^{\prime}\right)+c\right]\sin\theta dt\wedge d\theta \wedge d\phi.
\end{align}
Therefore, the corresponding chemical potential is given as
\begin{align}\label{eqapp22}
\psi_{\alpha}=-4\pi\left(\dfrac{2f^{\prime 2}(6+r_{+}f^{\prime})}{r_{+}}+c\right).
\end{align}
The second term in the parentheses is a pure gauge, which can be fixed by different methods.
\section{Coefficients of continued fraction expansion}\label{sec:Appendix}

Here, we present coefficients of continued fraction expansion up to fourth order. 
\begin{align}
&\epsilon=\dfrac{2M}{r_{+}}-1,\,\,\,\, a_{1}=-1-a_{0}+2\epsilon+r_{+}f_{1},\,\,\,\, a_{2}=-{\dfrac {4a_{1}-5\epsilon+1+3 a_{0}+ f_{2}r_{+}^{2}}{{ a_{1}}}},
\nonumber \\
&a_{3}=-\dfrac{1}{{a_{1}}{a_{2}}}[-{f_{3}}{{r_{+}}}^{3}+{a_{1}}{{a_{2}}}^{2}+5{a_{1}}{a_{2}}+6{a_{0}}+10{a_{1}}-9\epsilon+1]
\label{cfrac-a},\\
&a_{4}=-\dfrac{1}{{a_{1}}{a_{2}}{a_{3}}}\Big({f_{4}}{{r_{+}}}^{4}+{a_{1}}{{a_{2}}}^{3}+2{a_{1}}{{a_{2}}}^{2}{a_{3}}+{a_{1}}{a_{2}}{{a_{3}}}^{2}+6{a_{1}}{{a_{2}}}^{2}+6{a_{1}}{a_{2}}{a_{3}}+15{a_{1}}{a_{2}}+\nonumber\\
&10{a_{0}}+20{a_{1}} -14\epsilon+1\Big), \nonumber
\end{align}
and 
\begin{small}
\begin{align}
f_{4}=&\dfrac{1}{3r_{+}^{10}(r_{+}^{3}+48\alpha f_{1})(96\alpha f_{1}+r_{+}^{3}+48\alpha r_{+}f_{1}^{2})}\Big(-1248\alpha f_{1}r_{+}^{9}-7524\alpha f_{1}^{2}r_{+}^{10}+5971968\alpha^{5}f_{1}^{8}\nonumber\\
&+11064\alpha r_{+}^{11}f_{1}^{3}-5640192r_{+}^3\alpha^4 f_{1}^{7}-888192r_{+}^5\alpha^3 f_{1}^{5}+1769472r_{+}^{6}\alpha^3 f_{1}^{6}+1534464r_{+}^2\alpha^4f_{1}^{6}\nonumber\\
&-10944r_{+}^{7}\alpha^2 f_{1}^{3}+89136r_{+}^{8}\alpha^2f_{1}^4+4608r_{+}^{6}\alpha^2f_{1}^2-184896r_{+}^{9}\alpha^2f_{1}^{5}+86400\alpha^3 r_{+}^4f_{1}^{4}-344\times\nonumber\\
&f_{1}r_{+}^{13}-1344\alpha r_{+}^{10}\Lambda-1728\alpha r_{+}^{6}q^2+672\alpha q^4r_{+}^4+192\alpha q^6 r_{+}^2+192\alpha r_{+}^{14}\Lambda^3+576\alpha r_{+}^6\Lambda q^4-\nonumber\\
&44q^2r_{+}^{10}+960\alpha r_{+}^{8}\Lambda q^2-504576 r_{+}^6\alpha^3f_{1}^4\Lambda-504576 r_{+}^2\alpha^3f_{1}^{4}q^2 +116352 r_{+}^5\alpha^2 f_{1}^3 q^2+ 124416 \nonumber\\
&r_{+}^9\alpha^2f_{1}^3\Lambda+497664r_{+}^8\alpha^3f_{1}^4\Lambda^2-2985984r_{+}^4\alpha^4f_{1}^6\Lambda+1880064 r_{+}^3\alpha^3q^2f_{1}^5+1880064r_{+}^7\alpha^3\Lambda f_{1}^5\nonumber\\
&+864\alpha r_{+}^8+40320r_{+}^{10}\alpha^2f_{1}^{2}\Lambda^2-27648r_{+}^{12}\alpha^2f_{1}^{2}\Lambda^3-156672r_{+}^{11}\alpha^2f_{1}^3\Lambda^2-294912 r_{+}^{10}\alpha^2f_{1}^{4}\Lambda\nonumber\\
&-22\Lambda r_{+}^{14}-294912r_{+}^6\alpha^2f_{1}^4-27648\alpha^2f_{1}^2q^6+3936r_{+}^{13}\alpha f_{1}\Lambda^2-3360r_{+}^{11}\alpha f_{1}\Lambda-3552r_{+}^{7}\alpha f_{1}q^2\nonumber\\
&+5280 r_{+}^5 \alpha f_{1} q^4-2985984\alpha^4 f_{1}^6 q^2+497664\alpha^3 f_{1}^4q^4+12648 r_{+}^{12}\alpha f_{1}^2\Lambda +15192 r_{+}^{8}\alpha f_{1}^2q^2-\alpha^2 \nonumber\\
&16128r_{+}^{8}f_{1}^2\Lambda+31r_{+}^{12}-18432\alpha^2 r_{+}^4f_{1}^2q^2-82944r_{+}^4\alpha^2f_{1}^2\Lambda q^4-313344 r_{+}^7\alpha^2 f_{1}^3\Lambda q^2-82944r_{+}^8\nonumber\\
&\alpha^2f_{1}^2\Lambda^2q^2+9216r_{+}^9\alpha f_{1}\Lambda q^2+40320r_{+}^2\alpha^2q^4f_{1}^2+995328r_{+}^{4}\alpha^3f_{1}^4\Lambda q^2+80640r_{+}^6\alpha^2f_{1}^2\Lambda q^2+\nonumber\\
&156672r_{+}^3\alpha^2f_{1}^3q^4+576\alpha r_{+}^{10}\Lambda^2 q^2+288\alpha r_{+}^{12}\Lambda^2\Big).
\end{align}
\end{small}

\end{document}